**Artificial Intelligence and Big Data in Entrepreneurship: A New Era Has Begun**

Martin Obschonka*

*Australian Centre for Entrepreneurship Research, QUT, Australia*

David B. Audretsch

*Indiana University, USA*

April 19, 2019

* Corresponding author: Martin Obschonka, martin.obschonka@qut.edu.au

**Abstract**

While the disruptive potential of artificial intelligence (AI) and Big Data has been receiving growing attention and concern in a variety of research and application fields over the last few years, it has not received much scrutiny in contemporary entrepreneurship research so far. Here we present some reflections and a collection of papers on the role of AI and Big Data for this emerging area in the study and application of entrepreneurship research. While being mindful of the potentially overwhelming nature of the rapid progress in machine intelligence and other Big Data technologies for contemporary structures in entrepreneurship research, we put an emphasis on the *reciprocity* of the co-evolving fields of entrepreneurship *research* and *practice*. How can AI and Big Data contribute to a productive transformation of the research field and the real-world phenomena (e.g., "smart entrepreneurship")? We also discuss, however, ethical issues as well as challenges around a potential contradiction between entrepreneurial uncertainty and rule-driven AI rationality. The editorial gives researchers and practitioners orientation and showcases avenues and examples for concrete research in this field. At the same time, however, it is not unlikely that we will encounter unforeseeable and currently inexplicable developments in the field soon. We call on entrepreneurship scholars, educators, and practitioners to proactively prepare for future scenarios.

Keywords:

Entrepreneurship; Artificial Intelligence; AI; Big Data; Machine Learning; Smart Entrepreneurship

JEL classification: L26, M13, B41

# 1 Introduction

This essay and editorial for the Special Issue on artificial intelligence (AI), Big Data, and entrepreneurship is not an attempt to provide a complete overview and prospect of the scope and potentially disruptive nature of AI and Big Data for entrepreneurship research and practice. There is of course the "danger" that what we write and hypothesize here is incomplete or becomes obsolete very soon for various reasons. For example, any predictions and explications of future scenarios might quickly become outdated due to the rapid progress in the fields of AI and Big Data and their potentially far-reaching, yet hard to precisely predict, future implications for the real world (Grace et al. 2018). But there is of course also ongoing change with respect to the *nature* of entrepreneurship as a real-world phenomenon influenced by changing and evolving external enablers and institutions (Davidsson 2016; Davidsson et al. 2018; Eesley et al. 2016), and new generations of entrepreneurs who grew up in the digital era, the "digital natives" (OECD 2017).

Nevertheless, since there is growing attention around AI and Big Data in an increasing number of research and application fields, including industry, innovation, and business management, but comparably little attention with regard to entrepreneurship, we would like to offer some reflections, thereby being well aware of the challenging and uncertain nature of this endeavor, particularly with respect to long-term predictions within such dynamic fields. We write this piece from the perspective of entrepreneurship research, and we will try to focus on the *near* future, for example with respect to research priorities, infrastructure changes, and collaborative efforts in the coming generation of entrepreneurship research. Or as Alan Turing, one of the fathers of AI, once famously put it: "We can only see a short distance ahead, but we can see plenty there that needs to be done" (Turing 1950: 460).

AI and Big Data are growing in importance in broader research fields that are often seen as foundational for entrepreneurship research, such as economics (Acemoglu and Restrepo 2018; Brynjolfsson et al. 2017; Einav and Levin 2014), economic policy (Agrawal et al. 2019), innovation (Aghion et al. 2017), management (George et al. 2014; Ransbotham et al. 2017), and psychology (Kosinski et al. 2016). Likewise, this data revolution is also disrupting application fields associated with entrepreneurship, including industry, business management, and innovation (Cockburn et al. 2018). These massive transformations are also coined "the second machine age", in contrast to "the first machine age"—the Industrial Revolution 200 years ago (Brynjolfsson and McAfee 2014). So, this "second machine age" is driven not by coal and steam but by data and AI. Likewise, AI expert Andrew Ng describes AI as some kind of "new electricity", transforming industry and business in a fundamental way like electricity did 100 years ago (Burgess 2018). AI is therefore also affecting the concrete role and position *of people* in this new AI-infused world of work in many ways (Acemoglu and Restrepo 2018; Frank et al. 2019).

How are AI and Big Data defined? While there are different and evolving definitions available, AI can be broadly defined as intelligence demonstrated by machines—or, in terms of an academic field (typically seen as a subdiscipline of computer science), the examination of how digital computers and algorithms perform tasks and solve complex problems that would normally require (or exceed) human intelligence, reasoning, and prediction power needed to adapt to changing circumstances. This modern definition has been evolving since the first definition of AI presented by computer scientist John McCarthy more than 60 years ago as "the science and engineering of making intelligent machines" (see Andersen 2002). Within the AI terminology, an often-used categorization is that machine learning is a subset of AI, and deep learning (e.g., deep neural networks) a subset of machine learning. For a broad, general overview of AI and its techniques an often-recommended work is the textbook by Russell and Norvig (2016). Other overview works that focus particularly on machine learning are Bishop (2006) and Witten et al. (2016).

Big Data, in turn, which as an academic term is typically more loosely defined than AI, can mean a large volume of structured, semi-structured, or unstructured data, and a way to collect/produce, process, and analyze these datasets using non-traditional methods (e.g., AI methods). Such data are the fuel, "the new oil" (Agrawal et al. 2018) for AI and intelligent machines. Hence, Big Data and AI are often intertwined and go hand in hand as drivers of the current digital transformation in society (Brynjolfsson and McAfee 2014; Zomaya and Sakr 2017). However, it is also notable that AI can also rely on, and work with, smaller existing datasets (e.g., neural networks based on smaller amounts of existing data).

At first glance, the field of entrepreneurship research seems not well-prepared and probably overwhelmed by the rapid changes and progress in the field of AI and Big Data. After all, entrepreneurship is not really a subfield of computer science and vice versa. But, interestingly enough, when thinking about the topic of AI/Big Data in the context of entrepreneurship it leads us to essential questions such as rationality and uncertainty, intelligence and logic, discovery and ideation, strategy, and experienced-based pattern recognition—topics that have been of central concern for entrepreneurship researchers for many years. Moreover, one should probably go a step further by, somewhat radically, declaring that the era of AI and Big Data in entrepreneurship has clearly and inevitably begun, and this is true for both entrepreneurship *research* and *practice*. Intelligent machines and algorithms will not only inspire and empower a new generation of research in this field, they will also shape the actual real-world phenomenon of entrepreneurship, be it for example the development/recognition of opportunities and business ideas, "smart" entrepreneurial strategies, or the nexus and reciprocal interaction between entrepreneurial people and machine intelligence. We do not only assume that AI and Big Data will disrupt both entrepreneurship research and practice, it might also disrupt the way research and practice interact with each other. For example, it might change the way how research insights will be applied in the real world, and how entrepreneurial activity as a real-world phenomenon will inform future research. Finally, it might also help reducing *the distance* between entrepreneurship research and practice (e.g., the research field might apply knowledge and techniques from the practice field more quickly and directly, and vice versa).

We acknowledge though that these are somewhat bold predictions and that only the future can tell us whether we will see more change than stability at some point. In particular, it will be interesting to observe whether AI and Big Data will lead to a transformation of entrepreneurship research and practice in terms of more or less incremental or radical innovations (e.g., by "carrying out new combinations" of elements previously unconnected; Schumpeter 1934: 66) or whether it will indeed drive a complete paradigm shift and an "explosion of knowledge" that change the rule of the game (Kuhn 1970).

Notwithstanding these essential questions, in the following we would like to "risk" some concrete suggestions and reflections for both entrepreneurship research and practice, thereby considering the academic challenge of reflecting on entrepreneurship in terms of a research domain *and* a real-world phenomenon which are interlinked and co-evolving (Davidsson 2016).

Finally, we would like to offer a preliminary conclusion and outlook, before we introduce the Special Issue papers as concrete examples of already existing research projects in this field.

## 2 Research priorities

The consideration of AI and Big Data in entrepreneurship research is of course not a new phenomenon. However, the field could see *a new level* of AI-infused research projects. What are emerging/potential links between AI/Big Data and the diverse, interdisciplinary field of entrepreneurship research (Audretsch 2012; Shane 2012)? Arguably, the data revolution might bring unprecedented opportunities for entrepreneurship research and practice, but also new challenges and open questions. The central opportunity-aspect could be the promise of a new generation of insights into entrepreneurial phenomena revealed by means of new research methods, datasets, and study designs. The challenges and open questions, in turn, could range from ethical aspects and issues of privacy protection (D. Boyd and Crawford 2012), new statistical thinking and computational methods (Fan et al. 2014), to an adaptation and transformation of the whole research process (e.g., with respect to an open research culture that would not only ensure ethical standards but also transparency and reproduction of entrepreneurship research; Nosek et al. 2015). Therefore, both more conceptual and empirical work on AI, Big Data, and entrepreneurship is needed in future research.

*Conceptual work* could address the following and other issues:

1) Potential for productive vs. destructive entrepreneurship (Baumol 1990).

2) In what way does entrepreneurship of the future require the virtue of intelligence (Sternberg 2004) and the virtue of prediction power (Agrawal et al. 2018)?

3) Ways how AI/Big Data can support expert performance in the context of entrepreneurship (e.g., via supporting deliberate practice; Ericsson et al. 2018).

4) A practical introduction into relevant methods, technologies, and necessary hardware for conducting entrepreneurship studies utilizing Big Data and AI (e.g., computerized methods and specific software, smartphone methods, data mining, machine learning; Gosling and Mason 2015; Kosinski et al. 2016; Zomaya and Sakr 2017).

5) Issues of prediction vs. explanation; inductive, data-driven approaches vs. deduction, theory-driven approaches; and bigness vs. representativeness (Mahmoodi et al. 2017).

6) Potential dangers and ethical dilemmas associated with Big Data and AI methods in entrepreneurship research (e.g., with respect to data protection and privacy issues, generalizability of results, and the intrusiveness of data collection methods; D. Boyd and Crawford 2012).

7) Computerized language analysis (e.g., to identify entrepreneurial personality characteristics or relevant emotional states; R. L. Boyd and Pennebaker 2017; Eichstaedt et al. 2015).

8) How to use digital footprints from social networks and other sources (e.g., Twitter, Facebook, Instagram; Kosinski et al. 2013).

9) A reflection on Big Data and AI from a perspective of (non-)rationality, uncertainty, and risk (Kahneman 2002) and how this relates to entrepreneurship.

10) How can Big Data and AI methods improve entrepreneurship education and training?

11) Educating and training entrepreneurship researchers/budding entrepreneurs in Big Data and AI methods.

*Empirical work*, in turn, could address the following and other issues:

1) New Big Data-based metrics of entrepreneurial activity and quality (Guzman and Stern 2016).

2) Empirical research that can unlock the full potential of social media and other digital footprints for entrepreneurship research (Kosinski et al. 2013).

3) Identifying and predicting entrepreneurial characteristics and performance outcomes of people, teams, and organizations (Chen et al. 2012).

4) Formal and informal institutions of entrepreneurial regions (Glaeser et al. 2016; Obschonka 2017).

5) Entrepreneurship policy (Audretsch et al. 2007).

6) Entrepreneurial education and training (Fayolle 2007).

7) Networks (Wang et al. 2017).

8) Entrepreneurial finance (e.g., crowdfunding, analyses of investors and investment, and selection processes of high-potential startup projects; Block et al. 2018).

9) Simulation studies (Shim & Kim, 2018).

10) Analyses of populations that are underrepresented in existing entrepreneurship research (e.g., populations from outside of Western, educated, industrialized, rich, and democratic countries (WEIRD) or superstar entrepreneurs or entrepreneurial personalities in political leadership; Obschonka et al. 2017; Obschonka and Fisch 2017).

11) Business model processes relevant for entrepreneurial organizations and growth (Chen et al. 2017; Garbuio and Lin 2019; Hartmann et al. 2016).

12) Stress processes, well-being, health, and social behavior of entrepreneurial people and teams (e.g., using smartphone methods; Harari et al. 2017; Uy et al. 2010).

13) The study of biological factors and processes in entrepreneurship (e.g., genetics; Nicolaou et al. 2008).

14) Ecological sustainable entrepreneurship and the ecological/energy costs of AI and Big Data in entrepreneurial practice (Zeng 2017).

15) Potential future long-term effects on society and people of an AI-driven entrepreneurial economy (e.g., similar to research studying the long-term effects of the Industrial Revolution and today's cultural and psychological landscapes; Obschonka et al. 2018).

## 3 Potential impact of AI and Big Data on entrepreneurship as a real-world phenomenon

AI and Big Data might not only enrich and transform future entrepreneurship research, they might also transform at least some aspects of the actual real-world phenomena that entrepreneurship researchers usually study when they try to understand determinants and effects of the entrepreneurial process. Similar historical trends were observed in the recent past, such as the rise of e-commerce or new entrepreneurial finance "players" such as crowdfunding, which also influenced the agenda of entrepreneurship research (Block et al. 2018; Shepherd et al. 2019). So in other words, AI and Big Data might not only influence the methods but also the target that is studied with these methods in entrepreneurship research. As such, the research domain and the real-world phenomenon might co-evolve, or even co-transform, towards an AI-infused conglomerate of research and practice. This could mean that AI might further close the gap between entrepreneurship research and practice, with knowledge spillovers from research to practice and vice versa (or even blurring boundaries between research and practice).

How can AI and Big Data change the way entrepreneurship "happens" (e.g., how startups and intrapreneurial projects fail or succeed)? What implications are associated with these developments?

## 3.1 External enablers

First, such technological change can of course function as an *external enabler* for new entrepreneurial activity (Davidsson et al. 2018). It might actually be a prime example of radical external changes that empower and "enable" new economic activity that introduces new products and services via entrepreneurial means (for example AI startups like AIBrain, Banjo, or DeepMind), particularly when large incumbent firms struggle to fully embrace new technologies (Christensen 1997). This can also involve social entrepreneurship projects like, for example, OpenAI (cofounded by Elon Musk and not only motivated by the enabling power of AI but also by the associated, potentially existential risks for humanity).

On the other hand, startups in this field, particularly if they are faced with general resources constraints (including, for example, access to Big Data and powerful cloud computing), might also be confronted, challenged, and, to some degree, overwhelmed by the sheer pace of technological progress that transforms society and its institutions, and by the technological and knowhow requirements that need to be continuously updated to be at the forefront of this development. From this perspective, larger firms and organizations using AI and Big Data for intrapreneurial strategies could often enjoy a certain incumbent advantage over startups in that they might have better "economic muscles" and access to infrastructure and knowhow capital to participate in, benefit from, and even drive the ongoing data revolution (examples are Amazon Web Services (AWS) or Google.ai).

3.2 AI and Big Data plus humans

Second, whereas external enablers refer to the more or less *objective* external opportunity structure, AI and Big Data might also shape the way entrepreneurial *individuals and teams* participate in the entrepreneurial process. This leads us of course to the somewhat provocative question of whether AI and Big Data might be able to replace entrepreneurial people someday, or whether entrepreneurs and the new technologies will work hand in hand and form some sort of symbiosis.

The extreme scenario, where machines replace entrepreneurs as the central agent in the entrepreneurial process, would indeed lead to a complete paradigm shift. Arguably, this seems like an unlikely utopia though (at least for the near future; Brynjolfsson and Mcafee 2017), but similar discussions are currently held with respect to a large variety of jobs and human performance domains (Grace et al. 2018). If this utopia nevertheless comes to reality one day, it will disrupt the individual–entrepreneurship nexus (Shane 2003) and might basically erase human agency and subjectivity from the entrepreneurship equation, and thus probably also related research and practice fields such as the psychology of entrepreneurship or entrepreneurship education and training for humans. Moreover, maybe in the future one cannot avoid the fundamental question of why mankind would leave something as essential as entrepreneurship (e.g., discovering and analyzing opportunities, starting and growing an innovative business) to human agency, human bias, and human intelligence, and not (also) to powerful AI that might be able to outperform humans. Similar questions are being already (indirectly) asked around the topics of inventions, innovation, and scientific progress in general (Aghion et al. 2017; Cockburn et al. 2018; Slezak 1989). For example, recently Springer Nature published its first machine-generate scientific book (Writer, 2019). Just like AI has already started to massively shift the way in which research and practice is done in various fields such as astrophysics, genetics, and health, it could also shift the way entrepreneurship is carried out. However, it seems very unlikely today that AI will completely replace humans as entrepreneurial agents in the near future.

Interestingly, researchers estimate that jobs like business management (which might also involve entrepreneurship) are probably confronted with a relatively *small* likelihood to be replaced by machines in the near future because "generalist occupations requiring knowledge of human heuristics, and specialist occupations involving the development of novel ideas and artifacts, are the least susceptible to computerization" (Frey and Osborne 2017: 266). Indeed, "prototypical" entrepreneurship is often described as some sort of generalist work (Lazear 2004; Stuetzer et al. 2013) that involves human heuristics (Busenitz and Barney 1997), but it can of course also be linked to the more specialized development of novel ideas and artifacts (e.g., in the processes of ideation, invention, and innovation; Drucker 1985; Schumpeter 1934).

In the less extreme scenario, and for the near future the obviously more likely case, AI and Big Data do not replace but *support* entrepreneurs in performing entrepreneurial tasks and reaching their individual and organizational goals. So it would be less a matter of humans *against* machines than humans *plus* machines. This symbiotic, collaborative view seems to be evolving into the mainstream practice perspective in the contemporary business and management literature (Jarrahi 2018; Wilson and Daugherty 2018) and it should be a promising avenue for entrepreneurship as well.

Indeed, AI and Big Data have been supporting business activities and managers (e.g., finance, marketing, distribution, business planning, information systems, production, and operations) for a long time (Rajab and Sharma 2018), but the way and scope in which machines and humans can interact and collaborate in this field might be new. Wilson and Daugherty (2018), for example, describe five elements of business process improvement where AI and humans can collaborate: flexibility, speed, scale, decision-making, and personalization. The competitive advantage of mastering the new generation of AI and Big Data led Brynjolfsson and Mcafee (2017) to conclude that "over the next decade, AI won't replace managers, but managers who use AI will replace those who don't." Something similar could apply to entrepreneurs as well (e.g., "smart entrepreneurship"), particularly if they are operating in AI-relevant business fields or with ambitions where AI can help them to scale.

However, we see at least three fundamental questions associated with such a functional nexus between AI and entrepreneurship, and we can only touch the surface of these questions here. The first question concerns the very notion of *intelligence* itself. If AI is indeed a manifestation of intelligence where the latter is defined in a sense of *human* intelligence (Turing 1950), we can ask whether entrepreneurship really benefits from extremely high levels of human intelligence. Interestingly, so far research has not shown a clear link between (extremely high) intelligence and entrepreneurship. Intelligence researcher Robert J. Sternberg hypothesized that "successful entrepreneurship requires a blend of analytical, creative, and practical aspects of intelligence" (Sternberg 2004: 186). Hence, successful entrepreneurship might *not* be "a story about intelligence in the traditional sense" (e.g., general human intelligence; Spearman 1904) but rather about certain facets of intelligence that help entrepreneurs in their analytic, creative, and practical capacities. Hence, machine intelligence aiming to support entrepreneurship could focus on these domains, but we also have to stress that we clearly need more research on the role of both human and machine intelligence in entrepreneurship (and its changing nature) to come to safer conclusions. This could also include aspects of social and emotional intelligence (e.g., Bainbridge et al. 1994). However, others argue that it might be less intelligence and instead prediction that is the real promise of AI for business (Agrawal et al. 2018).

The second fundamental question concerns *uncertainty*. Entrepreneurship, as least in its classic sense, is often discussed as a context of uncertainty (McMullen and Shepherd 2006; Parker 2009) and hence as decision-making under the condition of uncertainty. Per this definition, and in contrast to the condition of risk for example, the concept of uncertainty implies that there are no reliable and complete preexisting data available that would help give orientation with respect to entrepreneurial tasks and causal models that come with a proven track record of success (Sarasvathy 2001). If this is true though, and if AI basically relies on existing data to learn patterns and to make predictions, we can ask whether AI will ever overcome these classic uncertainty challenges in entrepreneurship. Assuming a unique role of intuition, imagination, creativity, heuristics, and even cognitive bias in entrepreneurship, we can ask whether the irrational and rule-breaking nature many entrepreneurs show (Baron 1998; Obschonka et al. 2013) can indeed be imitated, "enhanced," or replaced by intelligent machines that might rely more or less on rules, rational logics, and formalized thinking. It seems that AI is better suited to create a "synthetic homo economics" (Parkes and Wellman 2015) than a rule-breaking, intuitive, and creative entrepreneur. It might contribute to what science historian Lorraine Daston describes as new rationalization driven by algorithms (that replaces self-critical judgments of reason, decision-making, and strategic planning; e.g., Erickson et al. 2013). On the other hand, AI under conditions of uncertainty is a vibrant research field (e.g., Ghahramani 2015; Tversky and Kahnemann 1974;

for an overview see also see Russell and Norvig 2016) and we might see major developments in this particular field in the near future as well, including new implications for entrepreneurship as effective decision-making under the condition of uncertainty (as it is discussed in game theory and with relation to similar strategy fields like public policy and warfare; Cummings 2017; Von Neumann and Morgenstern 1944).

These developments could also concern concrete essential entrepreneurial tasks in the very early stage of the entrepreneurial process, namely ideation and opportunity discovery/co-creation. Can AI help discover and co-create opportunities for new economic activity? Can it find its own data and make its own experience in the real world to learn how to be successful in this field? Again, this seems more like an unreal utopia at this point (e.g., given the sheer complexity of information and dynamics at various levels that might be involved; and also in view of the fundamental ongoing debate and disagreement in entrepreneurship research about what opportunities actually are; Alvarez et al. 2017; Davidsson 2017; Foss and Klein 2017; Ramoglou and Tsang 2016). Nevertheless, also with respect to aspects of ideation and autonomous learning, the field of AI is making incredible scientific progress. For example, recently an AI program that trained itself via reinforcement learning, *without any human input and guidance*, outperformed human experts and other AI programs trained by human experts in highly complex tasks (the board game Go and chess) (Silver et al. 2017; Singh et al. 2017). Other AI programs are obviously able to help creating pieces of "real" art (Cohen 2018), and economists stress the great potential of AI for discovery and inventions (Aghion et al. 2017; Cockburn et al. 2018).

The third fundamental question concerns potential *limits and risks* associated with AI and Big Data. It will be essential for the future not only to deal with ethical challenges and to enable and regulate access to data and technological infrastructure (e.g., for budding entrepreneurs but also entrepreneurship researchers), but also to maintain a critical perspective. For example, researchers raise the issue that AI might not always lead to the most reliable and useful solutions (e.g., Clever Hans effect; Lapuschkin et al. 2019) and it is important to ensure transparency and critical evaluation of AI performance. For example, in health research, scholars warn that "one of the dangers of ready accessibility of health care data and computational tools for data analysis is that the process of data mining can become uncoupled from the scientific process of clinical interpretation, understanding the provenance of the data, and external validation" (Belgrave et al. 2017). Indeed, there seems to be general danger of a "blind trust" in algorithms (Logg et al. 2019). So, if entrepreneurs rely on, but do not question and critically evaluate, AI results that might *not* actually deliver the best and smartest solution, it could lead to biased decisions and negative consequences or at least unused potential for the individual business.

## 3.2 Entrepreneurship education and training

Third and finally, besides their role as objective external enablers and as support for entrepreneurial individuals and teams, AI and Big Data might also enrich the practice field of *entrepreneurship education and training* in a new way. In an extreme case, where algorithms and intelligent machines overtake (some) entrepreneurial tasks from humans, this could mean, somewhat ironically, that these algorithms and machines themselves "receive" some sort of entrepreneurship education and training (e.g., in the form of real-world data that help them to learn). On the other hand, educators might want to make use of AI and Big Data to enhance their educational practices in the classroom and in other settings (McArthur et al. 2005). This could of course also mean actually teaching entrepreneurship-relevant AI/Big Data techniques to better prepare future entrepreneurs for this new era (e.g., to critically evaluate and interpret AI results). This could involve technological and knowhow aspects, but also ethical, social, and ecological issues. If AI and Big Data will indeed change the way entrepreneurs think and behave, the fields of education and training should be at the forefront of these developments. This could also involve other structures in society that are facilitating an entrepreneurial culture (Audretsch 2007).

**4 A preliminary conclusion and outlook**

We hope that, despite all the challenges, entrepreneurship researchers will initiate new, and participate in already ongoing conversations around the potential and concrete applications of AI and Big Data. The editorial of *Nature Machine Intelligence*, a recently founded major scientific journal that is devoted to AI and Big Data but also seeks: "to stimulate collaborations between different disciplines," concludes: "The pursuit of intelligent machines will continue to inspire in many ways, providing us with insights into human intelligence as well as stimulating technological and scientific innovation that could lead to future societal transformations. Now is the time to be part of the conversation" (*Nature Machine Intelligence* 2019: 1). We believe entrepreneurship research can and should be part of that broader movement. This could entail both intensified collaborations *with other* disciplines (e.g., computer science and information systems) and revised research structures *within* entrepreneurship research, be it new types of infrastructure (e.g., new research centers and access to the latest and powerful technologies and knowhow) or of communication channels and social structure (e.g., new types of conferences, workshops, data depositories and sharing possibilities, and maybe even academic journals; Landström and Harirchi 2018).

Finally, to stay within the opportunity-focused terminology of entrepreneurship research as "the study of *sources* of opportunities; the *processes* of discovery, evaluation, and exploitation of opportunities; and the set of *individuals* who discover, evaluate, and exploit them" (Shane and Venkataraman 2000: 218), and to stress the optimistic "promise"-factor that Shane and Venkataraman (2000) championed, a next concrete step could be the focus on the sources, processes, and individuals associated with opportunities around machine-based intelligence for new entrepreneurship research and its application. This at least could be one way to achieve an *entrepreneurial* handling of the dawning AI and Big Data-augmented era that seems to be in front of us.

## 5 The articles in this Special Issue

The central goal of the Special Issue "Rethinking the entrepreneurial (research) process: Opportunities and challenges of Big Data and Artificial Intelligence for entrepreneurship research" was to provide an interdisciplinary platform for conceptual and empirical papers addressing either opportunities or challenges (or both) of AI and Big Data for the diverse field of entrepreneurship research. Here we present seven articles that showcase examples for concrete research in this field. The works illustrate the usefulness of innovative perspectives and methods, but also the existing challenges and open questions that entrepreneurship researchers are facing when applying AI/Big Data to their field. All papers underwent a traditional double-blind peer review process.[1]

---

[1] We are grateful to all external reviewers, listed in the following, who reviewed papers that were submitted to this Special Issue project: Thomas H. Allison, Joern Block, Niels Bosma, Ryan Boyd, Dermot Breslin, Per Davidsson, Dimo Dimov, Uwe Dullek, Brent Clark, Graciela Corral de Zubielqui, Christian Fisch, Denise Elaine Fletcher, Michael Fritsch, Samuel D. Gosling, Sven Heidenreich, Colin Jones, Michal Kosinski, Edward J. Malecki, Alex Maritz, Nicos Nicolaou, Mark D. Packard, Luke Pittaway, Stuart Read, Enrico Santarelli, Dean Shepherd, Rolf Sternberg, Michael Stuetzer, Amulya Tata, Diemo Urbig, Frederik von Briel, Johan Wiklund, and Michael Wyrwich.

First, the articles by Coad and Srhoj (in press) and Obschonka et al. (in press) present empirical analyses that utilize AI and other Big Data techniques *to retest phenomena and mechanisms* that have already been studied with traditional methods in prior entrepreneurship research. Coad and Srhoj's (in press) article targets the prediction of high-growth firms. By means of a Big Data technique they identify valid predictors from a relatively large set of potential candidate predictors. They also stress though that with 10% explanation power, the prediction of high-growth firms remains a challenging endeavor despite the novel method they utilized. The article by Obschonka et al. (in press) focuses on entrepreneurial regions and attempts to measure and validate regional differences in entrepreneurial personality by utilizing individual-level large datasets and AI methods that extract psychological patterns. Specifically, they utilize large datasets collected from social media and AI-based language-analyses of this social media language from the various regions. They show that the AI-based measure of regional entrepreneurial personality that is solely based on freely available social media data is a similarly valid predictor and indicator of actual entrepreneurial activity in the region to regional personality measures collected from traditional self-reports (e.g., from millions of personality tests).

Second, the articles by Liebregts et al. (in press) and Zhang and Van Burg (in press) present *conceptual papers* that advance our theoretical thinking and knowledge about methods and data sources in the context of AI/Big Data and entrepreneurship. Liebregts et al. (in press) focuses on entrepreneurial decision-making and how behavioral and non-behavioral cues can be utilized via Big Data and AI. Zhang and Van Burg (in press) in turn address the question of how a subfield of AI—genetic algorithms—can be utilized in entrepreneurship when combining a design science and effectuation perspective.

Finally, the empirical articles by Kaminski and Hopp (in press), Prüfer and Prüfer (in press), and von Bloh et al. (in press) use AI/Big Data techniques *to examine relatively new research questions*. Kaminski and Hopp (in press) analyze crowdfunding campaign data (text, speech, and videos) using a neural network and language processing. Their novel analyses deliver interesting new implications for successful crowdfunding campaigns. The article by Prüfer and Prüfer (in press) examines demand dynamics for entrepreneurial skills by analyzing 7.7 million data points collected from job vacancies. They show which entrepreneurial skills are particularly important for which type of profession and also consider digital skills. von Bloh et al. (in press) present a novel analysis of news coverage of entrepreneurship and regional entrepreneurial activity, thereby addressing the potential link between media and entrepreneurship both from a conceptual and empirical perspective.